\documentclass[twocolumn,aps,prapplied]{revtex4-2} 

\usepackage{xcolor}
\usepackage{lineno}
\usepackage{graphicx}
\usepackage{soul}

\begin{document}

\title{Purcell-enhanced solid-state laser cooling}

\author{Mohammed Benzaouia}\email{Currently at School of Science and Engineering, Al Akhawayn University, Ifrane, Morocco.}\affiliation{Ginzton Laboratory and Department of Electrical Engineering, Stanford University, Stanford, 94305, California, USA}
\author{Shanhui Fan}\email{shanhui@stanford.edu}\affiliation{Ginzton Laboratory and Department of Electrical Engineering, Stanford University, Stanford, 94305, California, USA}

\begin{abstract}
We show that Purcell effect can lead to a substantial enhancement in the maximum cooling power for solid-state laser cooling. We numerically demonstrate such enhancement in a patterned slot-waveguide structure using ytterbium-doped silica as the active material. The enhancement arises primarily from the increase of saturation power density and the escape efficiency, and can persist in spite of the presence of parasitic absorption in the structure surrounding the active material. Our results point to a new opportunity in photonic structure design for optical refrigeration.
\end{abstract}

\maketitle

\section{Introduction}

Solid-state laser cooling allows the refrigeration of solids through anti-Stokes fluorescence using an external pump light~\cite{epstein2010optical, seletskiy2016laser}. While it was first suggested by Pringsheim in 1929~\cite{pringsheim1929zwei} and put on solid thermodynamic footing by Landau in 1946~\cite{landau1946thermodynamics}, the first experimental demonstration was not achieved until 1995 using a ytterbium (Yb) doped glass~\cite{epstein1995observation}. Experimental realization requires achieving both near unity external quantum efficiency and low parasitic absorption, which was successfully achieved in rare-earth doped glasses or crystals. More recent advances include cooling Yb$^{3+}$-doped YLiF$_4$ (YLF:Yb) crystal to a remarkable temperature of 87 K, making laser cooling the first solid-state refrigerator to reach cryogenic temperatures~\cite{melgaard2016solid, volpi2019optical}, as well as optical refrigeration of an infrared sensor payload to less than 135 K~\cite{hehlen2018first}. Laser cooling of levitated nanocrystals was also demonstrated~\cite{rahman2017laser}. Similar nanocrystals were used to refrigerate optomechanical resonators~\cite{pant2020solid}, water and physiological electrolytes~\cite{roder2015laser}, and electron-transparent silicon-nitride windows for transmission electron microscopy~\cite{dobretsova2021hydrothermal}. Solid-state laser cooling was also proposed as a promising method for the thermal management of relativistic lightsails~\cite{jin2022laser}. The recent demonstration of radiation-balanced ytterbium-doped silica fiber amplifier opens new exciting opportunities for athermal lasers~\cite{knall2021radiation}. 

Among the challenges that need to be addressed to further improve optical refrigeration is the saturation of the ions absorption which limits the maximum achievable cooling power, as noted in previous works~\cite{seletskiy2012cryogenic, lei2022laser, kock2022optical}. In this paper, we show that the saturation intensity, and therefore the cooling power, can be significantly increased using Purcell effect. Making use of the dependence of the saturation intensity on the decay rate, and the possibility of improving the decay rate by designing the photonic environment, as clarified in the pioneering work of Purcell~\cite{purcell1995spontaneous} and extensively explored in various nanophotonics structures~\cite{gerard1999strong, boroditsky1999spontaneous, vesseur2010broadband, poddubny2012microscopic, jacob2012broadband, akselrod2014probing}, we present a nanostructure design leading to a cooling power enhancement of $\sim 40$ compared to a bare active layer. This effect adds a new degree of freedom to be exploited, in conjunction with other physical (such as photonic, mechanical and thermal) considerations, when designing structures for laser cooling for different applications.  


\section{Theory}

\begin{figure}
\centering
\includegraphics[width=\linewidth,keepaspectratio]{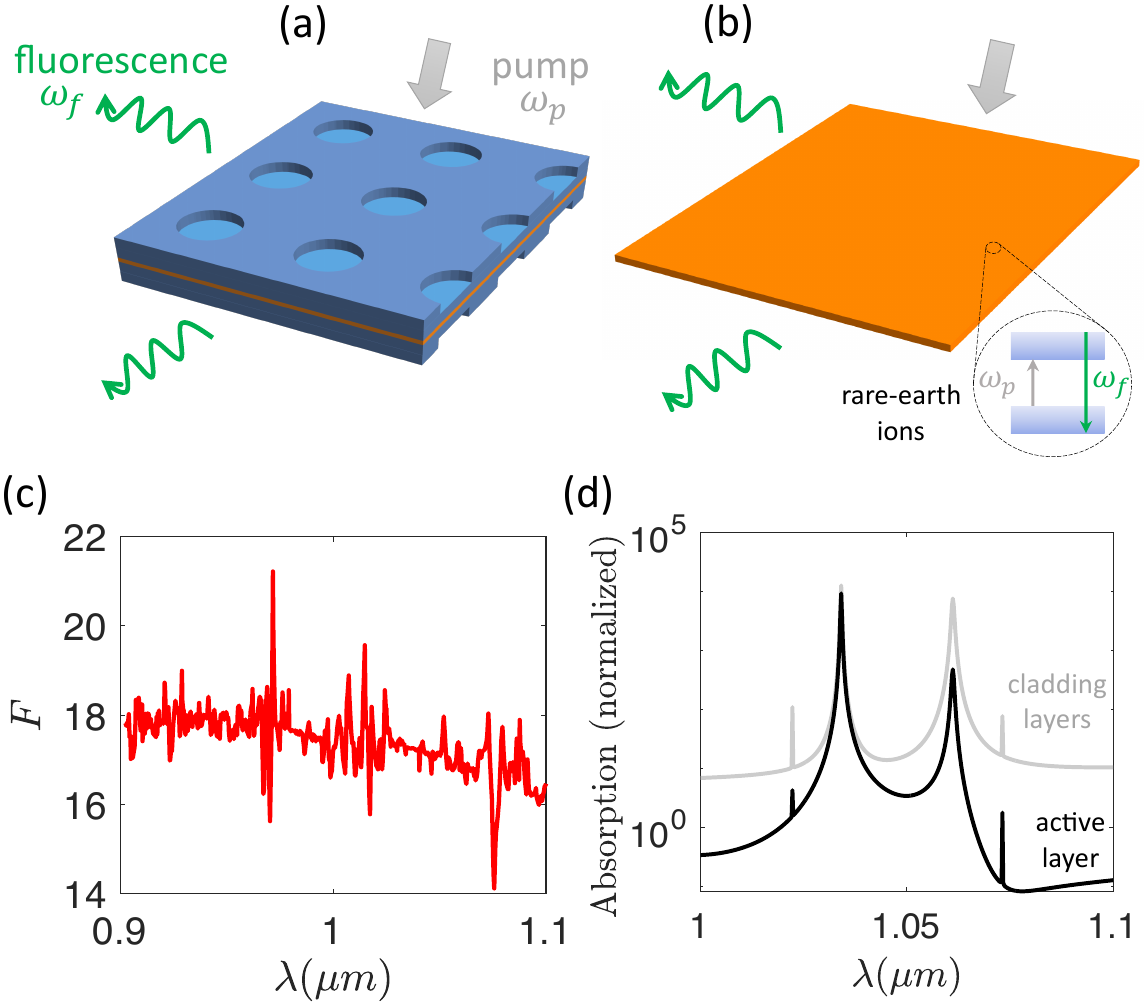} 
\caption{Enhanced laser cooling of a patterned slot waveguide nanostructure (a) as compared to a bare active layer (b). The active layer (ytterbium-doped silica with $10nm$ thickness) is surrounded by a patterned higher-index cladding layer (with refractive index 4 and thickness $86nm$). The surface pattern is a periodic square lattice of air holes with period $0.91\mu m$, diameter $0.39 \mu m$, and depth $13nm$ on each side of the structure. (c) Purcell factor $F$ defined as ratio of emitted power compared to a homogenous medium. (d) Absorption spectrum in both the active layer (black curve) and cladding layers (grey curve) normalized to the bare active slab absorption. All absorption coefficients are assumed to be equal in this plot.}
\label{Fig1}
\end{figure}

Laser cooling of solids has been successfully achieved in a variety of rare-earth doped materials. In these materials the cooling is achieved by upconversion of incident laser light in the fluorescent process (Fig.~\ref{Fig1}b, inset). The key factors for cooling efficiency are the mean fluorescence frequency $\omega_f$ and the external quantum efficiency $\eta_e=E\gamma_r/(E\gamma_r+\gamma_{nr})$ with $\gamma_r$ and $\gamma_{nr}$ being respectively the radiative and non-radiative decay rates and $E$ the fluorescence escape efficiency. For a given ions density $N_0$, the absorption coefficient $\alpha_r$ is equal to $N_0\sigma_a$ with $\sigma_a$ the ion absorption cross section. In addition to the absorption from rare-earth ions, there is an additional background absorption $\alpha_b$ from impurities and other materials which contributes to heating. For high pump powers, rare-earth ions absorption saturates at a saturation intensity $I_s$. On the other hand, since background absorbers are typically broadband and have low quantum efficiency, they can be assumed to be virtually unsaturable~\cite{sheik2019optimum, knall2021radiation}. The net cooling power per unit volume is then given by
\begin{equation}\label{eq:Pc-general}
p_c = \eta_c \frac{\alpha_r I}{1+I/I_s} - \alpha_b I, \; \text{with} \; \eta_c=\eta_e\frac{\omega_f}{\omega_p}-1,
\end{equation}
for a pump frequency $\omega_p$  and a local pump intensity $I$. The total cooling power $\bar{P}_c$ can be computed by spatially integrating Eq.~\ref{eq:Pc-general} over the structure. The system's temperature $T$ depends on the thermal load in the setup and satisfies $\bar{P}_c(T)=\bar{P}_{\text{load}}(T, T_0)$, with $T_0$ temperature of the surrounding. The minimum achievable temperature is reached when the thermal load is minimized. Here, we look at the improvement of the cooling power, regardless of the thermal load specific to the setup.

Due to absorption saturation, the cooling power reaches a maximum for an optimal pump intensity. From Eq.~\ref{eq:Pc-general} we find
\begin{equation}\label{eq:maxPc}
p_{c,max} = \left(\sqrt{\eta_c\alpha_r}-\sqrt{\alpha_b}\right)^2 I_s, \; \text{for} \; I / I_s =   \sqrt{\eta_c\alpha_r/\alpha_b}-1.
\end{equation}
$p_{c,max}$ can be improved by increasing $\eta_c\alpha_r$ and decreasing $\alpha_b$~\cite{tonkaev2019optical, ju2024purcell, mendicino2025purcell}. This is the same strategy needed to improve the cooling power also at small pump intensities~\cite{khurgin2007surface, melgaard2016solid}. However, $p_{c,max}$ can also be further improved by increasing $I_s$. For a two-manifold system, where the transitions relevant for cooling occur between a lower ground-state and an upper excited-state manifold as is the case of many rare-earth-doped materials, the saturation intensity is given by~\cite{volpi2021open}
\begin{equation}
I_s = \frac{\hbar \omega_p\gamma}{\sigma_a}\beta, \; \text{with} \;\beta = \frac{\sigma_a}{\sigma_a+\sigma_e},
\end{equation}
with $\sigma_e$ the emission cross section and $\gamma = E\gamma_r+\gamma_{nr}$ the extrinsic total decay rate. For a given zero-phonon transition energy $\hbar \omega_0$ (defined as the energy difference between the lowest levels of the lower and upper manifolds), $\beta^{-1}$ can be found using McCumber theory~\cite{mccumber1964einstein} as equal to $1+\exp\left[\frac{\hbar (\omega_0-\omega_p)}{kT}\right]$~\cite{peng2008temperature, mobini2019spectroscopic}. (For simplicity, we assumed that the partition functions of the upper and lower manifolds are equal.) Crucially, we note that $I_s$ can be increased by increasing the decay rate $\gamma$. This can be achieved using Purcell effect. In the limit of negligible parasitic absorption and near-unity quantum efficiency, and for a given bandwidth-averaged Purcell Factor $\overline{F}$, the maximum cooling power density becomes $p_{c,max} \approx N_0 \eta_c \hbar \omega_p \beta \gamma \overline{F}$. This result can be understood as following: each photon emitted by an ion extracts on average an energy equal to $\eta_c\hbar\omega_p\beta$ from the thermal bath. Therefore, by increasing the rate of emitted photons, cooling power is improved.   

\section{Structure}

To further explore the Purcell-enhanced radiative cooling effect, we consider the slot structure in Fig.~\ref{Fig1}a. It consists of an active layer and a surrounding higher-index cladding with periodic surface patterning. And to quantify the enhancement due to the slot structure, we will compare its performance to a bare active slab without the cladding (Fig.~\ref{Fig1}b). As demonstrated in previous works~\cite{jun2009broadband, yu2010fundamental, kalinic2020all, ourmanuscript}, this slot structure can exhibit a large broadband Purcell factor $F$ over an extended area and also avoids the trapping of emitted fluorescent light. $F$ is maximized for a large refractive-index contrast and thin active layer. As an example of the active layer, we consider $\text{Yb}^{3+}$ doped silica, a commonly used material for which solid-state laser cooling, operating around $1\mu m$ wavelength, was recently demonstrated~\cite{knall2020laser, topper2023potential}. To obtain a large Purcell factor, a cladding with large refractive index is needed. A small absorption loss is also needed to minimize additional parasitic heating from the cladding layer. Choices of material include GaAs, GaP or $\text{MoS}_2$ which have a refractive index in the range of 3 to 4 and small loss at the free space wavelength of $1\mu m$~\cite{hsu2019thickness, ermolaev2020broadband, wilson2020integrated, papatryfonos2021refractive}, with exact values depending on manufacturing process. For demonstration, we assume a constant index of 4 and study the dependence of cooling performance on the absorption loss value. 

The Purcell factor $F(r,\omega)$ in general can be computed by placing a dipole source with the emission frequency $\omega$ at the position $r$, and computing its emitted power, normalized by the emission of the same dipole source placed in a homogenous active medium. One then spatially average $F(r, \omega)$ to obtain $F(\omega)$, as shown in Fig.~\ref{Fig1}c. In our calculation, we use the MESH code that provides a more direct and efficient calculation of such spatial averaging~\cite{chen2018mesh, ourmanuscript}. We see a substantial broadband Purcell enhancement with frequency-averaged value $\overline{F}\sim 18$. The enhancement is due to the slot waveguide effect that localizes the field inside the thin active region~\cite{almeida2004guiding, xu2004experimental}. The large-period patterning ensures that all emitted fluorescent light escapes and is not trapped ($E=1$)~\cite{fan1997high}. Sharp features in Fig.~\ref{Fig1}c represent the large number of resonances that couple the slot waveguide modes to free space~\cite{ourmanuscript}. On the other hand, for the bare active slab, we confirm that a fraction of light is trapped ($E \approx 0.5$) which reduces the external quantum efficiency and decay rate.

In addition to changing emission properties, the patterned structure also changes absorption properties. Since the cladding can add substantial parasitic heating and cancel improvement from Purcell effect, it is important to minimize absorption in the cladding and maximize it in the active layer. This can be achieved by exploiting the resonances of the structure and maximizing active layer absorption at wavelengths where cooling is also maximum. In fact, cooling occurs in a wavelength window where $\omega_p$ is small enough to enable a large $\eta_c$, but still large enough to obtain substantial ion absorption $\alpha_r(\omega_p)$. For $\text{Yb}^{3+}$ doped silica, this is realized in the $1$-$1.1\mu m$ wavelength region. In Fig.~\ref{Fig1}d, we compute the ratios of absorption in the active layer and the cladding layers normalized to the bare active slab absorption, assuming normal-incidence pump and that all layers have the same absorption coefficient. These ratios are independent from the absorption coefficient value in the linear regime. Fig.~\ref{Fig1}d shows that active layer absorption is indeed maximized, compared to the cladding absorption, at around $1.035\mu m$. The figure also shows a substantial absorption enhancement compared to the bare active slab ($\sim 10^4$), which leads to higher total cooling efficiency, as defined by cooling power divided by external incident pump.  


\section{Cooling Enhancement}

In order to compute the cooling power as expressed in Eq.~\ref{eq:Pc-general}, we obtain the local intensity $I(r) = \frac{\epsilon_0 c n}{2}|E(r)|^2$ from direct rigorous coupled wave analysis (RCWA) simulation for a normal-incidence pump~\cite{liu2012s4}. Due to Purcell effect, $I_s$, $\omega_f$ and $\eta_e$ are also potentially affected. For a given Purcell Factor $F$ and a given experimentally measured fluorescence intensity spectrum $S(\omega)$~\cite{topper2023potential} , the new radiative decay rate $\gamma'_r$, for an emitter at position $r$,  is
\begin{equation}
\gamma'_r(r) = \gamma_r \frac{\int F(r,\omega)S(\omega)/\omega\; d\omega}{\int S(\omega)/\omega\; d\omega} = \gamma_r \overline{F}(r).
\end{equation} 
In our structure $\overline{F}(r)$ does not depend strongly on $r$ in the active layer. The modified fluorescence frequency is similarly given by $\omega_f'(r) = \int F(r,\omega)S(\omega)d\omega / \int \frac{F(r,\omega)S(\omega)}{\omega}  d\omega$. Since the Purcell factor is slightly higher at shorter wavelengths, as shown in Fig.~\ref{Fig1}b, there is a small blue shift in $\omega_f'$ which leads to an additional slight improvement in the cooling power. In this system, nonradiative contributions are dominated by quenching effects from ion-ion energy transfer. Here we assume that the rate of ion-ion energy transfer scales linearly with the emission rate~\cite{andrew2000forster, ghenuche2014nanophotonic}, and therefore the internal quantum efficiency $\eta_q$ is unchanged by the photonic structure since $\gamma_r$ and $\gamma_{nr}$ change by the same ratio, and hence $\eta_q$ depends on the ions concentration $N_0$ as~\cite{knall2021radiation}
\begin{equation}\label{eq:eta}
\eta_q \equiv \frac{\gamma_r}{\gamma_r+\gamma_{nr}} = \frac{1}{1+\frac{9}{2\pi}(\frac{N_0}{N_C})^2},
\end{equation} 
where $N_C$ is a critical concentration. On the other hand, the external quantum efficiency $\eta_e = \eta_qE/(1-\eta_q(1-E))$ is enhanced by the use of the photonic structure since the escape efficiency is improved as discussed earlier. Taking into account the effects mentioned above, the saturation intensity is enhanced as $I'_s(r) = I_s \left[\eta_q\left(E\overline{F}(r)-1\right)+1 \right]$.

\begin{figure}
\centering
\includegraphics[width=\linewidth, keepaspectratio]{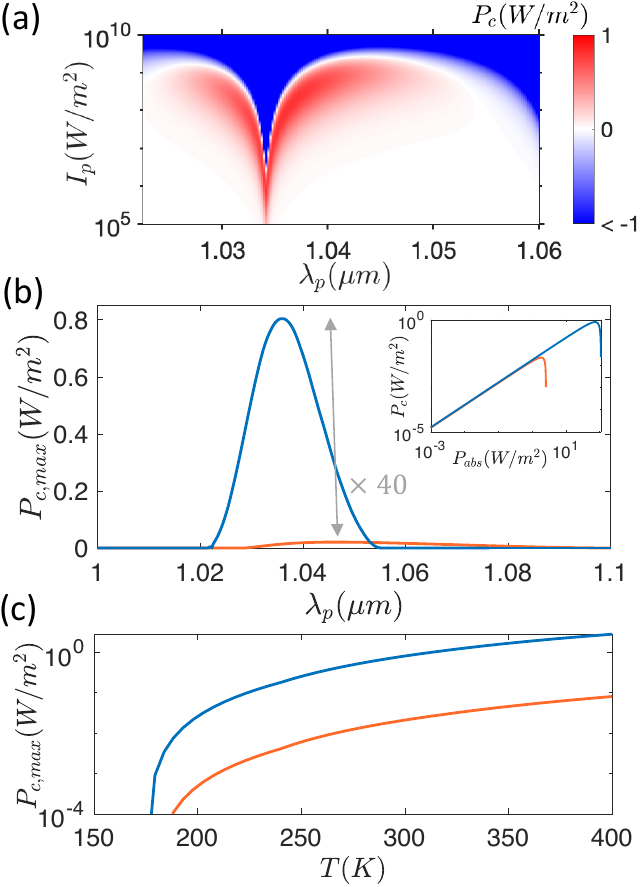} 
\caption{(a) Cooling power density $P_c $ spectrum as a function of pump wavelength $\lambda_p$ and intensity $I_p$ for the slot structure. (b,c) Maximum cooling power density $P_{c,max}$ as a function of the pump wavelength and temperature for the slot structure (blue curve) and for the bare active layer (orange curve). Inset shows saturation of $P_c$ as a function of the total absorbed power $P_{abs}$ at optimal pump wavelength ($\lambda_p \approx 1.036\mu m$ for the slot structure and $\lambda_p \approx 1.046 \mu m$ for the bare active layer). The cooling efficiency ($ = P_c/P_{abs}$) is around 2\% at low pump intensities. Only a small fraction ($\sim 10^{-4}$ at resonance for the slot structure) of incident power is absorbed.}
\label{Fig2}
\end{figure} 

\medskip
By integrating Eq.~\ref{eq:Pc-general} over a unit cell, we can now obtain the cooling power density as $P_c(\omega_p) = \int_{\text{unit cell}} p_c(\omega_p,r)dr/A_{\text{unit cell}}$. We use experimentally established parameters~\cite{topper2023potential} for $\sigma_a(\omega_p,T)$, $\omega_f(T)$ and $\gamma(T)$~\cite{mobini2019spectroscopic}. For the active layer, we take $\eta_q=0.99$, $N_0=6.56\cdot10^{25}m^{-3}$ and $\alpha_b=10\; dB/km \approx 2.3\cdot 10^{-3}m^{-1}$~\cite{topper2023potential}, and we also assume the same parasitic loss for the cladding layer for now. In Fig.~\ref{Fig2}, we show the cooling power density for different pump wavelengths $\lambda_p$, pump intensities $I_p$ and temperatures $T$. Figs.~\ref{Fig2}a,b correspond to room temperature. Due to patterning, cooling power has a resonant feature near $\lambda_p\approx 1.035\mu m$ as can be seen in Fig.~\ref{Fig2}a. For every pump wavelength $\lambda_p$, there is an optimal pump intensity $I_p$ at which the cooling power is maximum. The inset of Fig.~\ref{Fig2}b shows the cooling power density as a function of the absorbed power density, at the optimal pump wavelength for both the slot structure ($\lambda_p \approx 1.036 \mu m$) and the bare active layer ($\lambda_p \approx 1.046 \mu m$). For both structures there is an optimal absorbed power density that maximizes cooling.  For the slot structure, the optimum occurs at a much higher absorbed power density due to the increased saturation as induced by the Purcell effect.  In addition, the slot structure has an increased external quantum efficiency, and a small blue shift in $\omega'_f$, as compared with the bare active layer, which is beneficial for cooling, as well as parasitic heating in the cladding, which is detrimental. Combining all these factors, we see that the slot structure has a significantly enhanced cooling performance. We note that the primary contribution to the enhancement is from the increased saturation. In Fig.~\ref{Fig2}b, we plot the cooling power density at the optimal absorbed power density, for different pump wavelengths for both structures. The maximum cooling power density, as obtained by optimizing the pump wavelengths, is enhanced by a factor of $\sim 40$ in the slot structure. We also note that the pump wavelengths where the cooling is maximized are different due to the difference in the absorption spectra of these two structures.  

In Fig. \ref{Fig2}c, we plot the maximum cooling power density for the two structures as we vary the operating temperature. The slot structure shows a significantly improved maximum cooling power density for all temperatures. In addition, the slot structure also shows a modest improvement of the minimum achievable temperature (MAT), which can be beneficial for low-temperature applications of laser cooling. We note that for temperatures near MAT, the ions operate in the unsaturated regime, thus the improvement of MAT comes entirely from the improvement of the external quantum efficiency and the blue shift in the emission. However, as highlighted in other works~\cite{seletskiy2012cryogenic}, for practical applications, it is important to consider a saturation-limited MAT for a given desired cooling power. This saturation-limited MAT is substantially improved with Purcell effect as can be seen in the plot.

\begin{figure}
\centering
\includegraphics[width=\linewidth, keepaspectratio]{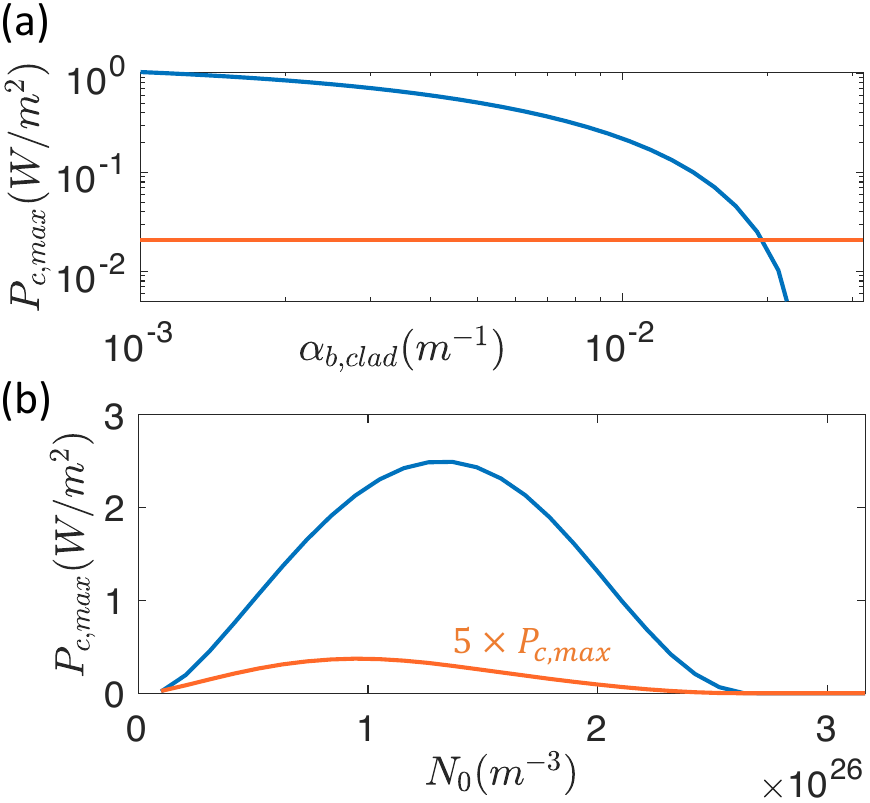} 
\caption{Maximum cooling power density (optimized over both pump intensity and pump wavelength) as a function of the cladding absorption $\alpha_{b,clad}$ (a) and rare-earth ion density $N_0$ (b) [orange curve in (b) is scaled $\times 5$ for clarity]. Blue curves correspond to the slot structure and orange curves correspond to the bare active layer.}
\label{Fig3}
\end{figure} 

\medskip
We further check the dependence of our results on different parameters. We first compute the maximum cooling power density as a function of the cladding's absorption coefficient as shown in Fig.~\ref{Fig3}a. As seen in the plot, the maximum cooling power density decreases with $\alpha_{b,clad}$, since the cladding loss only provides parasitic heating. In the limit of negligible cladding absorption, the enhancement can reach $\sim 65$. As $\alpha_{b,clad}$ is increased, the enhancement compared to the bare slab is still maintained for cladding loss up to $\sim 9\times \alpha_{b,active}$ (knowing that the thickness of each cladding layer is already almost 9 times larger than that of the active layer). In general, reducing the effect of parasitic absorption in the cladding requires focusing more incident pump inside the active layer and minimizing the overlap with the cladding. In applications where a cladding is needed for other purposes, including Purcell effect considerations in the design only improves the cooling power. We also check the performance as a function of the quantum efficiency, which is strongly influenced by quenching and therefore can be controlled by tuning the ions density $N_0$. In Fig.~\ref{Fig3}b, we compute $P_{c,max}$ for both structures as a function of $N_0$, using a critical quenching concentration $N_C \approx 1.61\cdot 10^{27} m^{-3}$~\cite{knall2021radiation}. We clearly see that the slot structure offers better cooling for a wide range of $N_0$. Two effects compete when increasing $N_0$: while the increase in $\alpha_r = N_0\sigma_a$ leads to a larger ion absorption, the decrease in $\eta_q$ (Eq.~\ref{eq:eta}) limits the cooling efficiency. In Fig.~\ref{Fig3}b, we indeed clearly see that the cooling power is limited for both large and small values of $N_0$.


\section{Conclusion}

We showed in this paper that Purcell effect can be used to substantially enhance solid-state laser cooling by increasing the corresponding saturation intensity and also improving the escape efficiency. As illustration we took Yb-doped silica, pumped at around $1\mu m$, but our results are applicable to different ions and hosts at different wavelength regions such as holmium or thulium ($\sim 2\mu m$). This gives the possibility of using different cladding materials (such as silicon or germanium) with high index and low loss, and can also help demonstrate laser cooling in new materials, particularly at larger wavelengths where the cooling efficiency is expected to improve. The use of multiple active layers, as suggested in Ref.~\cite{ourmanuscript}, can also be used to increase the total ions absorption. More generally, while we use a patterned-slot waveguide structure for demonstration, which ensures a large-area and broadband Purcell factor, other structures with substantially higher enhancement may also be explored in the future. Overall, the increase in cooling power due to Purcell effect, as demonstrated here, opens new opportunities for solid-state laser cooling design.

\bigskip
This work is supported by a grant from the U. S. Department of Energy (Grant No. DE-FG02-07ER46426), and by the Stanford Strategic Energy Research Consortium.

\bibliography{biblio}

\end{document}